\documentclass[a4paper,12pt]{article}
\usepackage{amsmath,amssymb,amsfonts,amsthm}
\usepackage{url}

\newcommand{\Rdm}{\mathbb{R}^2_{+}}

\newcommand{\rc}{^{(4)}\mathcal R}

\newcommand{\rt}{{^{(3)}R}}

\newcommand{\rd}{{^{(2)}R}}

\newcommand{\ckq}{\mathcal{L}_q}

\newcommand{\Lq}{\Delta_q}

\newcommand{\Ld}{\Delta}

\newcommand{\Ldt}{{^{(3)}\Delta}}

\title{Axially symmetric spacetimes: numerical and analytical perspectives}
\author{Sergio Dain\\
  Facultad de Matem\'atica, Astronom\'{i}a y F\'{i}sica, \\
     Universidad Nacional de C\'ordoba, \\
     Ciudad Universitaria, (5000) C\'ordoba, \\
  Argentina}

\begin{document}
\maketitle

\begin{abstract}
  Some new aspects of axially symmetric spacetimes are discussed. These results
  open the door for future interplay between analytical and numerical studies.
  The new developments are based on the role of the total mass in axial
  symmetry.  Finally, a list of relevant open problems is presented. These
  problems can be hopefully solved with an interaction between numerical and
  analytical insights.
\end{abstract}

\section{Introduction}\label{sec:introduction}
In any physical theory, the presence of a symmetry reduces the degrees of
freedom of the equations and hence it simplifies considerable the analysis.  To
study an isolated systems, the simplest models are the spherically symmetric
ones. However, it is well known that for vacuum Einstein equations, due to
Birkhoff's theorem, spherical symmetric spacetimes has no dynamics. The next
possible model with symmetries in vacuum are axially symmetric spacetimes. It
has been proved in \cite{bicak98} that no additional symmetry can be imposed to
the spacetime if we want to keep the gravitational radiation and a complete
null infinity. This result single out axially symmetric spacetimes as the only
models for vacuum isolated, dynamical, system with symmetries.  From this point
of view, axially symmetric gravitational waves are the simplest possible waves
emitted by isolated sources.

There exists many relevant physical models one can study in axial
symmetry. In particular, for vacuum and in the strong field regime, we list the
following  
\begin{itemize}
\item Head-on collisions of two black holes.

\item Rotating, non-stationary, black holes.  

\item Formation of black holes: weak cosmic censorship. 

\item Critical collapse of gravitational waves.

\end{itemize} 

What are the difficulties of axial symmetry?  To take advantage of the symmetry
an adapted coordinate should be used. However, the reduced equations are
formally singular at the axis.  This singular behavior introduce a major
difficulty in the analysis.  In fact, it can be argued that this singular
behavior near the axis is so complicated that the axially symmetric case is as
hard as the full general case.

What are the advantages of axial symmetry? The first obvious advantage for
numerical computations is that axially symmetric spacetimes are less
computationally expensive: only two effective spatial dimensions.  Also, the
number of equations and variables is reduced.  The second advantage, which is
also well known, is the conservation of angular momentum.  The angular momentum
is a quasilocal conserved quantity in axial symmetry (Komar integral of the
Killing vector).  That is, axially symmetric gravitational waves do not carry
angular momentum. In particular no Penrose process and no superradiant
scattering can occur for axially symmetric vacuum spacetimes.  This represents
an important simplification in the dynamics. The third advantage is the mass
integral formula. The total mass can be written as a positive definite volume
integral, as we will see. This is the main new ingredient that
we would like to discuss in the following.

The purpose of this review is to summarize some new results for axially
symmetric spacetimes with emphasis in the interplay between numerical and
analytical studies. The plan of the article is the following. In section
\ref{sec:axial-symmetry} we review the axially symmetric equations. The purpose
of this section is to present the relevant maximal-isothermal gauge and the
corresponding mass formula, which is valid only in this gauge. This formula
represents the main new motivation to study axial symmetry.  In section
\ref{sec:numer-relat-analyt} we discuss recent results, both numerical and
analytical which are based on the maximal-isothermal gauge and the mass
formula. Finally, in section \ref{sec:open-problems} we present relevant open
problems.

\section{Axial Symmetry}\label{sec:axial-symmetry}
Consider a vacuum solution of Einstein's equations, i.e., a four dimensional
manifold $M$ with metric $g_{\mu\nu}$ (with signature $(-+++)$) such that the
corresponding Ricci tensor vanishes
\begin{equation}
  \label{eq:132}
  \rc_{\mu\nu}=0. 
\end{equation}
Suppose, in addition, that the metric $g_{\mu\nu}$ admits  a
Killing field  $\eta^\mu$, that is $\eta^\mu$  satisfies the equation
\begin{equation}
  \label{eq:95}
  \hat \nabla_{(\mu} \eta_{\nu)} =0,
\end{equation}
where $\hat \nabla_\mu$ is the connection with respect to $g_{\mu\nu}$.  Greek
indices $\mu, \nu, \cdots $ denote  four dimensional indices. 

We define the square of the norm and the
twist of $\eta^\mu$, respectively, by
\begin{equation}
  \label{eq:1}
\eta =\eta^\mu\eta^\nu g_{\mu\nu}, \quad  \omega_\mu=\epsilon_{\mu\nu\lambda
  \gamma }\eta^\nu \hat
\nabla^\lambda\eta^\gamma.
\end{equation}
Using the field Eq. \eqref{eq:132}  it is possible to prove that 
\begin{equation}
  \label{eq:96}
  \hat \nabla_{[\mu} \omega_{\nu]}=0,
\end{equation}
and hence $\omega_\mu$ is locally the gradient of a
scalar field $\omega$
\begin{equation}
  \label{eq:2}
 \omega_\mu=\hat \nabla_\mu \omega. 
\end{equation}

Let $\mathcal{N}$ denote the collection of all trajectories of $\eta^\mu$, and
assume that it is a differential 3-manifold.
We define the metric $h_{\mu\nu}$  on $\mathcal{N}$ by
\begin{equation}
  \label{eq:3}
\eta g_{\mu\nu}=h_{\mu\nu}+ \eta_\mu\eta_\nu.
\end{equation}
 The vacuum field equations
\eqref{eq:132} can be written in the following form on $\mathcal{N}$
\begin{align}
  \label{eq:4}
 \Box \eta & =\frac{1}{\eta}(\nabla^a \eta \nabla_a \eta
- \nabla^a \omega \nabla_a \omega), \\
 \Box \omega & =\frac{2}{\eta}\nabla^a\omega \nabla_a \eta,\label{eq:4b} \\
\rt_{ab} &= \frac{1}{2\eta^2} (\nabla_a \eta\nabla_b \eta + \nabla_a \omega \nabla_b
\omega),  \label{eq:4c} 
\end{align}
where $\nabla_a$ and $\rt_{ab}$ are the connexion and the Ricci tensor of
$h_{ab}$, we have defined $\Box =\nabla_a \nabla^a$ and Latin indices
$a,b\ldots$ denote three dimensional indices on $\mathcal{N}$.

Up to this point, the only assumption we have made is that the spacetime admits
a Killing vector field $\eta^\mu$ and that $\eta^\mu$ is not null, otherwise 
the metric $h_{ab}$ is not defined. If the Killing field is timelike ($\eta<0$)
then the metric $h_{ab}$ is Riemannian and the equations
\eqref{eq:4}--\eqref{eq:4c} are the stationary Einstein vacuum equations.  On
the other hand, when the Killing vector is spacelike ($\eta>0$), the metric
$h_{ab}$ is a is a 3-dimensional Lorenzian metric (we chose the signature
$(-++)$). In axial symmetry, the Killing vector $\eta^\mu$ is spacelike and
its norm vanishes at the axis of symmetry. Hence, the equations are formally
singular at the axis. This singular behavior at the axis represents the main
difficulty to handle these equations.

In the Lorenzian case, Eq. \eqref{eq:4c} has the form of Einstein equations in
three dimensions coupled with effective matter sources produced by $\eta$ and
$\omega$. The effective matter Eqs. \eqref{eq:4}--\eqref{eq:4b} imply that the
energy-momentum tensor defined in terms of $\eta$ and $\omega$ by
\begin{equation}
  \label{eq:107}
  T_{ab}=\frac{1}{2\eta^2} (\nabla_a \eta\nabla_b \eta + \nabla_a \omega \nabla_b
\omega)-
\frac{1}{4\eta^2} h_{ab}(\nabla_c \eta\nabla^c \eta + \nabla_c \omega \nabla^c
\omega),
\end{equation}
is divergence free, i.e. $\nabla^aT_{ab}=0$. 

Eqs. (\ref{eq:4})--(\ref{eq:4c}) are purely geometric equations with respect to
the metric $h_{ab}$.  The essential point is that
there exist no dynamical degrees of freedom in 3-dimensional gravity (the Weyl
tensor vanishes) and hence all the dynamics is produced by the effective matter
sources determined by $(\eta,\omega)$. In other words, the dynamics of vacuum
axially symmetric gravitational waves has the behavior of matter moving in a
lower dimensional space. Hence, it can be expected that the total mass of the
waves can be computed as an ``integral on the matter sources''. As we will see
in the following, this is precisely what happens and it leads to the mass
formula in axial symmetry.

\subsection{2+1 decomposition}
\label{sec:2+1-decomposition}
In order to formulate an initial value problem, we will perform an standard
$2+1$ decomposition of Eqs. (\ref{eq:4})--(\ref{eq:4c}). Note that this is
completely analogous to the $3+1$ decomposition of Einstein equations, in fact
all the formulas are formally identical because the dimension does not appear
explicitly in them (see, for example, \cite{rendall-pdeigrgtim2008},
\cite{rendall08}).
  
Consider a foliation of spacelike, 2-dimensional slices $S$ of the metric
$h_{ab}$. Let $t$ be an associated time function  and let $n^a$ be
the unit normal vector orthogonal to $S$ with respect to the metric
$h_{ab}$. The intrinsic metric on $S$ is denoted by $q_{ab}$ and is given by
\begin{equation}
  \label{eq:6}
h_{ab}=-n_an_b + q_{ab}.
\end{equation}
Define the density $\mu$ by
\begin{equation}
  \label{eq:5}
\mu = 2\rt_{ab} n^a n^b+ \rt,
\end{equation}
 and the current $J_b$ by
\begin{equation}
  \label{eq:36}
  J_b=-q^c_bn^a \rt_{ca},
\end{equation}
where  $\rt=\rt_{ab}h^{ab}$  denotes the trace of $\rt_{ab}$.
 Then, using Eq. \eqref{eq:4c} we obtain
\begin{align}
  \label{eq:7}
\mu &= \frac{1}{2\eta^2}\left(\eta'^2 + \omega'^2 + |D\eta|^2 + |D\omega|^2
\right),\\
J_A &= -\frac{1}{2\eta^2}\left(\eta' D_A\eta  +  \omega' D_A\omega \right),
\end{align}
where $D_A$ is the connexion with respect to $q_{AB}$. The prime denotes
directional derivative with respect to $n^a$, that is
\begin{equation}
  \label{eq:14}
\eta' = n^a\nabla_a \eta= \frac{1}{\alpha} \left(\partial_t\eta- \beta^AD_A
  \eta\right) 
\end{equation}
where $\alpha$ is the lapse and $\beta^A$ is the shift vector of the
foliation.
The indices $A,B, \cdots$ denote two dimensional indices on $S$. 
The constraints equations corresponding to (\ref{eq:4c}) are  given by 
\begin{align}
  \label{eq:8}
\rd  -\chi^{AB}\chi_{AB}+\chi^2 &=\mu,\\
 D^A\chi_{AB} -D_B\chi &=J_B, \label{eq:8b}
\end{align}
where $\rd$ is the Ricci scalar of $q_{AB}$,  $\chi_{AB}$ is the
second fundamental form of $S$ and $\chi$ its trace
\begin{equation}
  \label{eq:133}
  \chi= q^{AB}\chi_{AB}. 
\end{equation}
We use the following sign convention for the definition of $\chi_{AB}$
\begin{equation}
  \label{eq:sce}
  \chi_{ab}= - q^c_a \nabla_c n_b= -\frac{1}{2}\pounds_n q_{ab},
\end{equation}
where  $\pounds$ denotes Lie derivative.
The  evolution equations are given by
\begin{align}
  \label{eq:127}
 \partial_t q_{AB} &= -2\alpha \chi_{AB} + \pounds_\beta q_{AB},\\ 
\label{eq:38}
\partial_t \chi_{AB} & = \pounds_\beta \chi_{AB}- D_AD_B\alpha +\alpha \tau_{AB},
\end{align}
where
\begin{equation}
  \label{eq:134}
  \tau_{AB}=\chi \chi_{AB}  + \rd_{AB}-\rt_{AB}-2\chi_{AC}\chi^{C}_{B}.
\end{equation}
and  
\begin{equation}
  \label{eq:136}
  \rt_{AB}=\frac{1}{2\eta^2} (\partial_A \eta \partial_B \eta + \partial_A
  \omega \partial_B \omega).  
\end{equation}
The evolution equations
\eqref{eq:127}--\eqref{eq:38} and the constraint equations
\eqref{eq:8}--\eqref{eq:8b} constitute a complete $2+1$ decomposition of the
3-dimensional Einstein Eq. \eqref{eq:4c}. It remains to decompose the
effective matter Eqs. \eqref{eq:4}--\eqref{eq:4b}. The result is the following
\begin{align}
  \label{eq:149}
  -\Sigma'' + \Lq \Sigma + D_A \Sigma \frac{D^A \alpha}{\alpha} + \Sigma' \chi
  & =\frac{1}{\eta^2} \left(\omega'^2 -|D\omega|^2 \right),\\
 \label{eq:149b}
-\omega'' + \Lq \omega + D_A \omega \frac{D^A \alpha}{\alpha} + \omega' \chi 
&= \frac{2}{\eta^2}\left(D_A\omega D^A\eta -\omega'\eta'\right),
\end{align}
where  $\Lq$ is the
Laplacian with respect to $q_{AB}$, i.e.  $\Lq=D^AD_A$ and we have defined
$\Sigma=\log \eta$. 

The evolution equations \eqref{eq:127}--\eqref{eq:38} and
\eqref{eq:149}--\eqref{eq:149b} together with the constraint equations
\eqref{eq:8}--\eqref{eq:8b} represent the complete $2+1$ decomposition of the
geometrical equations (\ref{eq:4})--(\ref{eq:4c}) in an arbitrary gauge.  This
set of equations presents an important feature which is peculiar of $2+1$
dimensions. Namely, the evolution equations \eqref{eq:127}--\eqref{eq:38} and
the constraint equations \eqref{eq:8}--\eqref{eq:8b} are, roughly speaking,
equivalent in the following sense.  With an appropriate gauge choice (in the
next section we will present an important example) the constraint equations
\eqref{eq:8}--\eqref{eq:8b} form an elliptic system that determines $q_{AB}$ an
$\chi_{AB}$ in terms of the effective matter sources $\eta$ and
$\omega$. Hence, the evolution equations \eqref{eq:149}--\eqref{eq:149b}
together with the constraint equations \eqref{eq:8}--\eqref{eq:8b} constitute a
complete system of equations. In particular, a solution of this system will
automatically satisfy the other evolution equations
\eqref{eq:127}--\eqref{eq:38}.  Alternative, it is possible to solve the
evolution equations \eqref{eq:149}--\eqref{eq:149b} and
\eqref{eq:127}--\eqref{eq:38}. If the initial data satisfy the constraint
equations at a given time, then the solution will satisfy the constraint
equations for all times. Note that even when the gauge is fixed, we have
different possibilities for constructing evolutions schemes. The first choice
is called in the numerical literature ``constrained evolution'' and the second
choice ``free evolution''. It is also possible to mix them to obtain ``mixed
evolution'' schemes (see the discussion in \cite{Rinne:thesis}).

\subsection{Gauge and mass formula}
\label{sec:gauge}
In this section we describe the maximal-isothermal gauge.  In particular we
review the mass formula valid in this gauge (see \cite{Dain:2008xr} for
details).  For the lapse, we impose the maximal condition on the 2-surfaces
\begin{equation}
  \label{eq:40}
  \chi=0.
\end{equation}
Note that we are not imposing that the surfaces are maximal in the
3-dimensional picture as in \cite{Dain:2008xr}. The later condition is the one
generally used (see, for example, \cite{Choptuik:2003as} \cite{Rinne:2008tk}),
but the difference is only minor. In particular the mass formula is positive
definite for both conditions. The one used here appears to be natural with
respect to the rescaled metric $h_{ab}$. Eq. \eqref{eq:40} implies the
following well known equation for the lapse
\begin{equation}
  \label{eq:37}
  \Lq \alpha = \alpha(\chi^{AB}\chi_{AB}+ \mu_1 ),
\end{equation}
where 
\begin{equation}
  \label{eq:45}
 \mu_1= \rt_{ab}n^an^b=  \frac{1}{2\eta^2}\left(\eta'^2 + \omega'^2 \right).
\end{equation}
The maximal gauge \eqref{eq:40} can be, of course, imposed in any dimensions
and  it is not related at all with axial symmetry. In contrast, the
condition for the shift is peculiar for two space dimensions.   
The shift vector is fixed by the requirement that 
the intrinsic metric $q_{AB}$ has the following form 
\begin{equation}
  \label{eq:10}
q_{AB}=e^{2u}\delta_{AB},
\end{equation}
where $\delta_{AB}$ is a fixed (i.e. $\partial_t \delta_{AB}=0$) flat metric in
two dimensions.  Then, using \eqref{eq:40}, we obtain that the trace free part
of \eqref{eq:127} is given by
\begin{equation}
  \label{eq:42}
  2\alpha\chi_{AB}=(\ckq \beta )_{AB},
\end{equation}
where $\ckq$ is the conformal Killing operator in two dimensions with respect
to the metric $q_{AB}$.  Equation
(\ref{eq:42}) is an elliptic first order system of equations for $\beta^A$.

 The elliptic Eqs. (\ref{eq:37}) and (\ref{eq:42}) determine lapse and
 shift for the metric $h_{ab}$ and hence they fix completely the gauge freedom in
 Eqs. \eqref{eq:4}--\eqref{eq:4c}. This gauge has associated a natural
 cylindrical coordinate system $(t, \rho, z)$ for which the metric
 $\delta_{AB}$ is given 
\begin{equation}
  \label{eq:66}
  \delta = d\rho^2+dz^2, 
\end{equation}  
and the axis of symmetry is given by $\rho=0$. The slices $S$ are the half
planes $\Rdm$. 

It is useful to introduce the following auxiliary functions. Instead of $\eta$ we will
use the function $\sigma$ defined by 
\begin{equation}
  \label{eq:72}
  \eta=\rho^2 e^\sigma.
\end{equation}
Note that for Minkowski $\eta=\rho^2$, that is, $\sigma$ measure the non-flat
part of the norm $\eta$. 
Also, instead of $u$ it is convenient to work with the function $q$ defined by 
\begin{equation}
  \label{eq:70}
  u=\log \rho + \sigma + q.
\end{equation}
The important property of $q$ is that it vanished at the axis. This is a
consequence of the regularity conditions of the metric at the axis (see
\cite{dain10} for details).

The equations presented in section \ref{sec:2+1-decomposition} can be
explicitly written as partial differential equations in the coordinates $(t,
\rho, z)$ (see \cite{dain10}).  In particular 
the Hamiltonian constraint is given by 
\begin{equation}
  \label{eq:101}
  \Ldt \sigma +\Ld q = -\frac{\epsilon}{4},
\end{equation}
where
\begin{equation}
  \label{eq:102}
 \epsilon =   \frac{e^{2u}}{\eta^2}\left(\eta'^2 + \omega'^2 \right)+ |\partial
 \sigma|^2 + \frac{|\partial \omega|^2}{\eta^2} +  2 e^{-2u}  \hat\chi^{AB} \chi_{AB}. 
\end{equation}
The hat on $\chi^{AB}$ means that the indices are moved with the flat metric
$\delta_{AB}$,  $\partial$ denotes partial derivatives with respect to $\rho$
and $z$, $\Ld$ is the 2-dimensional flat Laplacian, namely
\begin{equation}
  \label{eq:11}
 \Ld  =\partial^2_\rho+\partial^2_z,
\end{equation}
and $ \Ldt$ is the 3-dimensional flat Laplacian acting on axially symmetric
functions, that is
\begin{equation}
  \label{eq:12}
  \Ldt=\Ld +\frac{ \partial_\rho}{\rho}.
\end{equation}

The positive scalar $\epsilon$ plays the role of the energy density of the
gravitational waves in this gauge. The integral of $\epsilon$ over the slice
$S$ is given by  
\begin{equation}
  \label{eq:103}
  m= \frac{1}{16}\int_{\Rdm}   \epsilon\, \rho \,
  d\rho dz.   
\end{equation}
It is a non-trivial fact that this number $m$ is precisely the total ADM mass
of the spacetime. Moreover, the fall off properties of the solutions $\alpha$
and $\beta$ of gauge conditions
ensure that the mass is conserved along the evolution, namely 
\begin{equation}
  \label{eq:9}
  \frac{d}{dt} m =0,
\end{equation}
see \cite{Dain:2008xr} for details. 

The mass formula (\ref{eq:103}) together with the conservation law (\ref{eq:9})
represent a relevant property of this gauge which is not present in other ones.

In any physical theory conserved quantities (in particular, conserved energies)
are very important to control the evolution of the system. However, in General
Relativity, the conserved mass appears as a boundary integral and not as a
volume integral (as, for example, in the wave equation). Hence it is not
possible to relate the mass with any norm of the fields to control the
evolution of them (for the wave equation the energy is precisely the norm of
the wave). The mass formula (\ref{eq:103}) for axially symmetric systems in the
maximal-isothermal gauge represents a remarkable exception.

\section{Results}
\label{sec:numer-relat-analyt} 

\subsection{Numerical results}
\label{sec:numerical-results}

Axially symmetric spacetimes has been studied numerically since the very
beginnings of  numerical relativity, see chapter 10.4 in
\cite{alcubierre-it3nrsomop2008} and references therein for an historical
perspective. A very good reference for early results is the review article
\cite{Bardeen83}. In particular, in this article the maximal and isothermal
gauges are discussed. Another important review article is
\cite{Nakamura87}, where the symmetry reductions and $2+1$ decomposition
mentioned in section \ref{sec:axial-symmetry} was used for first time in
numerical relativity.

As we mentioned in the introduction, the difficulty introduced by the singular
behavior at the axis is severe and that made axially symmetric evolution
particularly difficult to handle. As a consequence of this ``axisymmetric codes
were practically abandoned as soon as computers became powerful enough to be
able to handle full three-dimensional simulations in the early 1990s.'' (the
quote is from \cite{alcubierre-it3nrsomop2008}).

New insights in the study of the axial regularity problem in numerical
relativity were presented in \cite{Garfinkle:2000hd}. They introduced two main
ingredients. The first one is that the behavior at the axis is treated as a
singular boundary condition which are handled by introducing ``ghost zones''
outside the axis. The second one is that all the variables used are such that
either they vanish at the axis or have vanishing derivative there, but not
both. This cure the axis instabilities from the numerical point of view.  

Later in \cite{Choptuik:2003as} a new code was developed which also use similar
techniques to handle the axis problem. This article is particularly relevant in
our context because they use essentially the same equations and the same gauge
conditions as presented in section \ref{sec:axial-symmetry}. However, the mass
formula (\ref{eq:103}) for this gauge was not known at that time and hence it
was not used in that article. 

Further studied were performed in \cite{Rinne:thesis} \cite{Rinne:2005sk},
where new evolutions schemes are proposed and the general case with twist is
considered. In \cite{Ruiz:2007rs} the regularization procedure at the axis was
performed for more general evolutions schemes.  In \cite{Rinne:2008tk} 
important issues concerning the maximal-isothermal gauge are studied, in
particular a new choice of variable is proposed in order to guarantee the
uniqueness of solutions of the relevant equations. I will come back to this in
the next section.

In a very recent series of articles, the critical collapse of gravitational
waves is studied \cite{Sorkin:2009wh} \cite{Sorkin:2010tm}. This code use a
different gauge choice, namely harmonic coordinates.  Finally, we mention
\cite{dain10} where the linear equations in the maximal-isothermal gauge were
studied and boundary conditions compatible with the mass formula were
presented.

\subsection{Analytical results}
\label{sec:analytical-results}

The first analytical application of the mass formula is the proof of the
following geometrical inequality
\begin{equation}
  \label{eq:8h}
 \sqrt{|J|} \leq  m,
\end{equation}
where $J$ is the total angular momentum of the spacetime (see \cite{Dain05e}
\cite{Dain06c} for details).  Extension of this theorem, which remove technical
assumptions and also simplify considerable its proof were presented in
\cite{Chrusciel:2007dd} \cite{Chrusciel:2007ak}.  The remarkable aspect of this
proof if that extreme Kerr black hole appears as the minimizer of the mass
integral.  This suggests stability properties of these black holes. In
particular, it suggests that it is perhaps easier to prove stability near an
extreme black holes than a non-extreme one in axial symmetry (see the essay
\cite{Dain:2007pk}). There exist important extensions to inequality
(\ref{eq:8h}) which include charges \cite{Costa:2009hn}
\cite{Chrusciel:2009ki}.

The second geometrical inequality proved with a suitable quasi-local formulation of
the mass formula is the following
\begin{equation}
  \label{eq:5h}
8\pi |J|\leq  A,
\end{equation}
where $A$ is the area and $J$ is the angular momentum of the black hole
horizon. See \cite{dain10d}, \cite{Acena:2010ws} \cite{Dain:2011pi} for
details. In the stationary case with matter and charge, it has been proved
in \cite{Hennig:2008zy}. Numerical evidences for the validity of this inequality
has been presented in \cite{Jaramillo:2007mi}. 

Inequality (\ref{eq:5h})  ensures that the Christodoulou quasi local-mass of the
black hole is a monotonically increasing quantity in the evolution (see the
discussion in \cite{dain10d}).

Both geometrical inequalities are proven on the initial data, that is, the
proofs do not make use of the evolution equations.  In this gauge the
equations reduce to a coupled hyperbolic--elliptic system which is formally
singular at the axis. Due to the rather peculiar properties of the system, the
local in time existence has proved to resist analysis by standard methods. To
analyze the principal part of the equations, which may represent the main
source of the difficulties, we study linear perturbation around the flat
Minkowski solution in this gauge. In \cite{Dain:2010ne} we solved this
linearized system explicitly in terms of integral transformations in a
remarkable simple form. This representation is well suited to obtain useful
estimates to apply in the non-linear case.

\section{Open Problems}
\label{sec:open-problems}
There exists two relevant open model problems for axially symmetric spacetimes:  

\begin{enumerate}
\item Black hole formation and critical collapse of axially symmetric
  gravitational waves (Numerical).

\item  The stability of the Kerr black hole in axial symmetry (Mathematical).

\end{enumerate}
These problems are expected to be difficult, they constitute the long term
motivation of this study.  In the following I will briefly discuss them in
order to propose simpler open problems that can contribute to the resolution of
1 and 2 and at the same time are feasible to solve in the near future.

Most of the numerical work mentioned in section \ref{sec:numerical-results} is
concerned with the vacuum equations and with the critical collapse of the
gravitational waves. As we pointed out in the introduction, there exists other
interesting physical systems in axial symmetry. For example, the
collapse of an axially symmetric rotating star studied in
\cite{Nakamura87}. However, if we restrict ourselves to the vacuum equations (as
we do in this article) then probably the most relevant open problem to study in
numerical axial symmetry is the critical collapse. And the reason is that this
problem can not be handled, with present computers, without symmetry
assumptions due to the high resolution needed.  The subject has already a long
history (see the review article \cite{lrr-2007-5}), and despite many efforts
the problem still lacks definitive answer (see the recent article
\cite{Sorkin:2010tm}).

Although there exist many possible gauge choices in axial symmetry, the mass
formula suggests the maximal-isothermal gauge as privileged one. Also,
constrained evolutions schemes appear to be better than free evolutions
schemes (see the discussion in \cite{Rinne:2008tk}). In the following we focus
on the maximal-isothermal gauge in its constrained formulation. In the article
\cite{Rinne:2008tk} it was clearly pointed out one of the main problems of
these equations. There are many elliptic equations and it is not clear a priori
that they will always admit a unique solution along the evolution.  A new set of
variables was proposed in that article to ensure that the equations have  the
correct behavior at the linearized level to ensure uniqueness of solution.
However, it is still not clear if the non-linear equations (notably, the
Hamiltonian constraint) is always solvable. It is an open problem to see if
these equations (or some variant of them) are well posed. I believe that this
is an important problem that stress the interplay between analytical and
numerical studies. Perhaps the resolution of the well-posedness question will
lead us to select (or even discover) the correct evolution scheme.

Another question which is important in numerical evolution is the issue of
outer boundary conditions on a finite domain. In the article \cite{dain10} a
set of boundary condition at the linear level were proposed. This boundary
conditions have the property that the energy appears to leave the domain, at
least for the class of initial data studied in this article.  It is important
to note that the mass formula provides a well-defined local measure of the
gravitational energy. These boundary conditions have not yet been implemented
for the complete equations.  

The following two problems summarize the discussion above.

\begin{itemize}  

\item[1.a.] \textbf{Well-posedness for the maximal isothermal gauge.} Is the Hamiltonian
  constraint always solvable? Does the gauge breaks down in the strong field
  regime?

\item[1.b.] \textbf{Implementation of the radiation boundary conditions for the
    full equations.}  How much they improve the evolution scheme?  It is
  possible to prove (may be with some extra assumptions) that for such boundary
  conditions the energy  always leaves the domain? 

\end{itemize}

As we have seen, problem 1 is more numerically oriented, since it appears to be
very difficult to understand  critical phenomena with the current analytical
tools.  In contrast problem 2 is more analytical. In recent years there have
been important new developments in the problem of the linear stability of the
Kerr black hole from the analytical point of view.  Most of this work is
concerned with scalar waves on a fixed Kerr black hole background. We refer to
the review articles \cite{Dafermos:2008en} and \cite{Dafermos:2010hd} and
reference therein. These works are not restricted to axial symmetry. The
expectation is, of course, that axially symmetric spacetimes are simpler. In
particular, axial symmetry provides extra geometrical structure that can give
new insight into the equations. As we mentioned in section
\ref{sec:analytical-results} much of the current evidences that this is the
case come from the geometrical inequalities (\ref{eq:8h}) and (\ref{eq:5h}).
These inequalities have been proved under some restricted conditions. In the
case of the quasi-local inequality (\ref{eq:5h}), the only restriction is that
the initial data are assumed to be maximal. It is important to remove this
restriction or to find counter examples. Numerics can play a relevant role in
the search of possible counter examples. Also, it would be very interesting to
generalize this inequality to include the charge. This is important not only
because the charge is a relevant parameter of a black hole but also because the
charge is well defined without any symmetry assumption (this is not the case of
the quasi-local angular momentum).  The natural question is whether a similar
inequality holds without axial symmetry. This would involve an appropriate
definition of quasi-local angular momentum. Another interesting problem is to
study what happens in the case of equality in (\ref{eq:5h}). A rigidity
property has been proved in \cite{Dain:2011pi} that characterize the local
geometry of the surface in that case. However, it is very likely that this kind
of surface can only appear as asymptotic limit as the  extreme Kerr
initial data.  A related question is to
study small deformation of the extreme Kerr initial data, as the ones
constructed in \cite{Dain:2010uh}. Do these initial conditions admit apparent
horizons if the perturbation is small enough? This question is relevant because
if the answer is negative then these initial conditions have the chance to
provide complete Cauchy surfaces that are outside the black hole region, as in the
extreme Kerr black hole.

Regarding the global inequality (\ref{eq:8h}), the most relevant open problems
is to prove it for multiple black holes and also to remove the maximal
condition. The multiple black holes case is related with the uniqueness of the
Kerr black hole among stationary black holes with disconnected horizon
components.

All the previous problems are concerned with the initial conditions. The
maximal-isothermal gauge will be useful to answer questions regarding the black
hole stability if the mass formula can be used in some way to control the
evolution.  A natural first problem would be to recover the non-linear
stability of Minkowski \cite{Christodoulou93} in this gauge. The expectation is
that the mass formula will provide a simpler (and different) kind of approach
to this problem; although, of course, always restricted to axial symmetry. A
second problem it is the study of axially symmetric perturbations of a black
hole in this gauge.  The expectation is that the mass formula can help to prove
linear stability under axially symmetric perturbations of the Kerr black hole.
The advantage of this problem in comparison with the non-linear stability of
Minkowski in this gauge is that it is probably easier to deal with and also it
will prove something new, since the linear stability of the Kerr black hole is
still an open question.

We summarize this discussion in the following list of open problems in
relation to problem 2.

\begin{itemize}  

\item[2.a.] \textbf{Geometrical inequalities.} Remove the maximal assumption. For the
  quasi-local inequality (\ref{eq:5h}) include the charge and study the
  possible generalization of this inequality without any symmetry
  assumptions. Study the case of the equality: it is possible to have such a
  surface in an asymptotically flat initial data? Study the existence of
  minimal surfaces or apparent horizon for small perturbation of extreme Kerr
  initial data. For the global inequality (\ref{eq:8h}), prove it for multiples
  black holes. 

\item[2.b.] \textbf{Linear stability of the Kerr black hole in axial symmetry.} Use
  the mass conservation to control the linear evolution.

\end{itemize}


\end{document}